%Paper: astro-ph/9503043
%From: mj@moon.astrouw.edu.pl (Michal Jaroszynski)
%Date: Fri, 10 Mar 1995 11:03:47 +0100 (MET)

%
% Plain TEX, no figures
%
% The full text with figures can be obtained as compressed postscript file by
%  anonymous ftp to astro.princeton.edu; cd library/preprints;
%  get pop615.ps.Z (compressed binary file, 75 kB)
%

\magnification=1200 \def\wc{\hangindent=4em \hangafter=1 \noindent}
\baselineskip 14pt \parskip 3pt \null 
\headline={\ifnum\pageno=1\hfil\else\hfil\tenrm--\ \folio\ --\hfil\fi}
\footline={\hfil}

\centerline{\bf DIFFRACTION EFFECTS IN MICROLENSING OF Q2237+0305}
\vskip 1.0cm
\centerline{M. Jaroszy{\'n}ski$^{1,2,3}$ and B. Paczy\'nski$^{1,4}$ }
\vskip 0.5cm

\centerline{ $^1$ Princeton University Observatory, Peyton Hall,
                Princeton, NJ 08544 USA}

\centerline{ $^2$ Warsaw University Observatory, Al. Ujazdowskie 4,
                00-478 Warsaw, Poland}

\centerline{$^3$ E-Mail: mj@astrouw.edu.pl}

\centerline{$^4$ E-Mail: bp@astro.princeton.edu}

\vskip 1.0cm
\centerline{\it Received: ...........................................}
\vskip 0.5cm
\centerline {\bf ABSTRACT}

Geometrical optics provides an excellent description for quasar images
crossing caustics which are formed by gravitational microlensing of
objects like Q2237+0305.   Within this approximation the source size
can be estimated from the maximum magnification reached at caustic crossings.
We evaluate the limitations imposed by diffraction on caustics
using the formalism developed by Ulmer \& Goodman (1995).
Close to a caustic a new characteristic length, smaller that the Fresnel
length, enters the problem, limiting the angular resolution to
about 0.2 pico arcsecond, or equivalently $ \sim 3 \times 10^{9}~{\rm cm}$
at the source.  To achieve this resolution the brightness must be
monitored at time intervals of a few seconds.  If a significant fraction
of quasar luminosity comes from sources smaller than those limits then
interference effects would make the observed intensity oscillate,
in a close analogy with a two slit experiment.  The characteristic
period of such oscillations is expected to be about one tenth of
a minute.  If such oscillations are detected then photometry carried
out at a single site may permit the determination of the caustic
transverse velocity, and therefore may permit a direct conversion of the
time units of brightness variations to the linear units at the source.

\wc{{\it Subject headings:} Gravitational lensing - dark matter
- quasars: structure -quasars: Q2237{\-}+0305  }

\vfill\eject

\vskip 0.5cm
\centerline {\bf 1. INTRODUCTION}
\vskip 0.5cm

The quadrupole gravitational lens 2237+0305 (Huchra et al. 1985) is the
first, and so far the best studied case of gravitational microlensing
at large optical depth.  The four macroimages are formed by the whole
galaxy acting as an astigmatic lens.  The four lines of sight pass through
four different inner regions of the lensing galaxy and are subject
to microlensing by the stars located near those lines of sight.
Some of the four macroimages were observed to
vary on a time scales of months (Racine 1992 and references therein)
and even days ( Pen et al. 1994).  The optical depth to macrolensing
combined with the shear make the magnification of all four
macroimages so large (cf. Wambsganss \& Paczy\'nski 1994,
and references therein) that the microlensing is due to a collective
effect of many stars, as emphasized by Wambsganss (1992).  This is
best seen at the magnification patterns, or equivalently the illumination
patterns which clearly
demonstrate that high magnification events are mostly caused by the
source crossing one of the many caustics created by the microlensing,
and almost none is caused by an isolated star
(Wambsganss, Paczy\'nski \& Schneider 1990).

The generic structure of almost every caustic crossing is the following.
On one side of the caustic the total magnification of a point source varies
as $ A = A_{\rm min}  + \left( d_{\rm c} / r_{\rm s} \right) ^{1/2} $
(cf. Fig. 1),
where $ A_{\rm min} $ is the magnification on the other side of the
caustic, $ r_{\rm s} $ is the linear distance between the source and the
caustic, and $ d_{\rm c} $ is a characteristic length related to the caustic.
In this paper all lengths, distances, and velocities in directions
perpendicular to the line of sight are measured in the lens plane, while
their equivalents in the source plane are denoted by the same symbols
with hats.
%%%%%%%%%%%%%%
The length scale $ d_{\rm c} $ is proportional
to the Einstein ring radius $ r_{\rm E} $
of individual microlenses, the stars in the lensing galaxy.
%%%%%%%%%%%%%%%%%
If the source moves
with a velocity $ V $ at an angle $ \theta $ to the caustic then
the magnification varies with time as
$$
A \approx A_{\rm min}
 + \left( d_{\rm c} / \Delta t V \sin \theta \right) ^{1/2}
= A_{\rm min} + \left( \Delta t_{\rm c} / \Delta t \right) ^{1/2}  ,
\eqno(1a)
$$
$$
\Delta t_{\rm c} = d_{\rm c} / V \sin \theta = \alpha r_{\rm E} / V \sin \theta
\sim M^{1/2} / V \sin \theta ,
{}~~~~~~~~ \alpha \equiv d_{\rm c} / r_{\rm E} ,
\eqno(1b)
$$
where $ \Delta t_{\rm c} $ is the characteristic time related to the caustic,
$ M $ is the mass of microlensing stars, and $ \alpha $ is a dimensionless
parameter of the order unity.  The magnification
of a point source increases all the way to infinity upon crossing any
caustic.  When the caustic is crossed the image magnification drops
discontinuously to a finite value $ A_{\rm min} $.
Naturally, the crossing is equally likely to be in the opposite direction,
with a discontinuous increase of the image magnification from
$ A_{\rm min} $ to infinity, followed by a gradual decline described with
the equations (1a,b).

If the source has a finite size, $ d_{\rm s} $, then the maximum magnification
at the caustic crossing is
$$
A_{\rm max} \approx A_{\rm min}
 + \left( d_{\rm c} / d_{\rm s} \right) ^{1/2} ,
\eqno(2)
$$
and the discontinuous jump in the magnification is replaced by a change
on a time scale
$$
\Delta t_{\rm s} \approx d_{\rm s} / V \sin \theta ,  \eqno(3)
$$
where $ \Delta t_{\rm s} $ is the characteristic time related to the source.

Note that the high magnification events caused by caustic crossings are
characterized by two distinctly different time scales.  One time scale,
$ \Delta t_{\rm c} $ is proportional to the square root of some average mass
of the stars that are responsible for microlensing.
The other time scale, $ \Delta t_{\rm s} $ is proportional to the source size.
With frequent and accurate photometric measurements it should be possible
to partly deconvolve the source structure (Grieger, Kayser, \& Refsdal 1988;
Wambsganss \& Paczy\'nski 1991).

Pen et al. (1994) presented the first observations of 2237+0305 variability
which were interpreted as caused by a caustic crossing.  The
image A increased its brightness by $ \sim 1.5 $ magnitude over two
months, corresponding to $ \Delta t_{\rm c} \approx 3 \times 10^6 ~{\rm s}$,
and subsequently its brightness dropped by $ \sim 1.5 $ magnitude in
less than a week, corresponding to
$ \Delta t_{\rm s} \approx 3 \times 10^5~{\rm s}$.
If our transverse velocity with respect to the lensing galaxy at
$ z_{\rm g} = 0.04 $ is $ \sim 10^3 ~{\rm km}~{\rm s}^{-1}$,
then the projected velocity
at the source at $ z_{\rm s} = 1.7 $ is
$ \hat{V} \sim 10^4~{\rm km}~{\rm s}^{-1}$,
and the source size may be estimated to be
$ \hat{d}_{\rm s} \sim \hat{V} \theta \Delta t_{\rm s} \sim 3 \times 10^{14} ~
cm $.  This is
just $ \sim 20 $ astronomical units, much less than the estimates based
on accretion disk models: $ d_{\rm s} \approx 1.5 \times 10^{16} ~{\rm cm} $
(Rauch \& Blandford 1991)
or $ \hat{d}_{\rm s} \approx 4 \times 10^{15} ~{\rm cm} $ (Jaroszy\'nski,
Wambsganss, \& Paczy\'nski 1992, cf. also
Czerny, Jaroszy\'nski, \& Czerny 1994; Jaroszy\'nski \& Marck 1994;
Witt \& Mao 1994).

There is a possible way to reconcile the results of Pen et al. (1994) with
the standard disk model of the quasar 2237+0305 if the event was caused by
a very fast moving caustic crossing the image of the quasar.  Such fast
caustics were found by Kundi\'c \& Wambsganss (1993) and by Kundi\'c, Witt
\& Chang (1993) to be the consequence of random motion of the stars
responsible for microlensing.  Only future observations and the well
sampled light curves are likely to resolve this issue.

In the geometrical optics approximation the caustics are infinitely sharp
and in principle the resolution of quasar structure is limited only by
the accuracy and frequency of photometric measurements during the
rapid change of the apparent brightness caused by a caustic crossing.
However, at some level diffraction effects must
limit the resolving power of any microlensing system.
The aim of this paper is to determine what is the highest resolution
that might be achieved.
In the next section we treat the problem theoretically and in Sec.3 we apply
our calculations to Q2237+0305. Discussion of the
observability of the effects follows in the last section.

\vfill
\eject

\centerline{\bf 2. DIFFRACTION BY CAUSTICS}
\vskip 0.5cm

The diffraction effects near optical caustic have the same nature as in a
two-slit experiment: they are caused by the interference of each photon
passing the two paths corresponding to the two images which appear or
disappear when the source crosses the caustic.
The interference of radiation coming to an observer along different paths
through the gravitational field, has been investigated by several authors.
(Schneider \& Schmidt-Burgk 1985; Deguchi \& Watson 1986; Peterson \& Falk
1991; Gould 1992; Stanek, Paczy{\'n}ski \& Goodman 1993; Ulmer \& Goodman
1995).
In most cases the calculations were limited to the simplest case of a single
point-mass lens.   The effects of physical optics in gravitational
lensing are referred to as femtolensing (Gould 1992).

Recently Ulmer \& Goodman (1995, hereafter UG)
have presented the method of calculating
femtolensing effects for a general gravitational lens system.
In particular they obtained the solution for the case of a source
close to a caustic.
In this paper we estimate the importance of the femtolensing for quasars.
We are interested in the effects detectable with a broad band photometry,
rather than the femtolensing effects in the spectra.

We use a general expression for the time delay along a ray crossing the
lens surface at point ${\bf r} \equiv (x,y)$ and coming from a source
point ${\bf r_{\rm s}} \equiv (x_{\rm s},y_{\rm s})$ as projected onto the lens
plane
(cf. Blandford \& Narayan 1986):
$$
c\tau(x,y,x_{\rm s},y_{\rm s})={1+z_{\rm g} \over 2D}\left( (x-x_{\rm
s})^2+(y-y_{\rm s})^2 \right)
		  +(1+z_{\rm g})c\tau_{\rm g}(x,y)
\eqno (4)
$$
where $\tau$ is the total time delay and $\tau_{\rm g}$  is the delay
caused by the gravitational field of the lensing galaxy,
$D \equiv D_{\rm g}D_{\rm gs}/D_{\rm s}$
(in standard notation) is the characteristic gravitational lens distance
and $z_{\rm g}$ is the redshift of the lens.
Suppose we choose our coordinate system in such a way, that locally,
near the origin, the line $x=0$ is tangent to a critical line related to
a fold caustic. In the simplest case we have $\tau_{,xx}=0$ on the
critical line $x=0$. For a system consisting of point lenses one has
$\tau_{{\rm g},xx}+\tau_{{\rm g},yy} \equiv 0$ everywhere,
except singular points
located at point masses. This equation fixes the value of the
$\tau_{,yy}$ derivative on the critical line, where $\tau_{,xx} =0$.
Introducing $\psi$ - a quantity proportional to the time delay and
expanding it in the lowest nontrivial order we have:
$$
\psi \equiv {Dc\tau \over 1+z_{\rm g}}
= {1 \over 6}a x^3+y^2-x_{\rm s}x-y_{\rm s}y+{1 \over 2}x_{\rm s}^2+{1 \over
2}y_{\rm s}^2
\eqno (5)
$$
where $a$ is defined by the third x-derivative of the time delay and can
be made positive by the transformation $x \rightarrow -x$.
There are other parametrizations possible near a critical line
(i.e. Schneider \& Weiss 1986;  Kayser \& Witt 1989; Witt 1990;
Witt, Kayser, \& Refsdal 1993). The caustic line
in our case is locally given by $x_{\rm s}=0$.
Using equation (1a) we have near the caustic:
$$
A-A_{\rm min}= \left( {d_{\rm c} \over r_{\rm s}} \right) ^{1/2} ,
\eqno (6)
$$
where $r_{\rm s}$  is a distance of the source from the caustic line and
$d_{\rm c}$ is a parameter characterizing the caustic.
In our case $r_{\rm s} = x_{\rm s}$. The magnification due to the caustic
related
images is proportional to the
inverse of the $||\psi_{,ij}||$ determinant, where $\psi$ is given by
equation (5). Comparing with the above formula we get:
$$
d_{\rm c} = {1 \over 8a}
\eqno (7)
$$
and substituting into the expression for $\psi$ we have:
$$
\psi = {1 \over 48d_{\rm c}}x^3+y^2-x_{\rm s}x-y_{\rm s}y
+{1 \over 2}x_{\rm s}^2+{1 \over 2}y_{\rm s}^2  .
\eqno (8)
$$

Physical optics calculations introduce the Fresnel length, which in our
case can be defined as:
$$
d_{\rm F}= \left( {\lambda D \over 2\pi} \right) ^{1/2}
\equiv \left( {c D \over \omega} \right) ^{1/2}  ~~,
\eqno (9)
$$
where $\lambda$ is the wavelength and $\omega$ the frequency of light.
With this definition one can calculate the complex amplitude (UG):
$$
\Psi(x_{\rm s},y_{\rm s})={-i \over 2\pi d_{\rm F}^2}
\int_{-\infty}^{+\infty}dx~\int_{-\infty}^{+\infty}dy
{}~\exp\left({i\psi \over d_{\rm F}^2}\right),
\eqno (10)
$$
The normalization is such, that in the absence of the lens
$\Psi(x_{\rm s},y_{\rm s})=1$.
In the case of a point source near the caustic line we get:
$$
|\Psi(x_{\rm s},y_{\rm s})|=
2^{1/3}~\pi^{1/2}~\left({d_{\rm c} \over d_{\rm F}}\right)^{1/3}
{}~|{\rm Ai}(-r_{\rm s}/d_{\rm f})|~~,
\eqno (11)
$$
where $\rm Ai$ is the Airy function (Abramovitz \& Stegun 1964)
and the characteristic scale for its argument changes is:
$$
d_{\rm f} =
{ d_{\rm F} \over 2^{4/3}}~\left({d_{\rm F} \over d_{\rm c}}\right)^{1/3} .
\eqno (12)
$$
Both results are in agreement with UG but we use different parametrization.
Using asymptotic expansion of the Airy function one can see, that for
$r_{\rm s} >> d_{\rm f} $ the flux from a point source
$F(r_{\rm s})=|\Psi|^2 \rightarrow \left( {d_{\rm c}/r_{\rm s}} \right) ^{1/2}
$
as in geometrical optics calculation.

We have checked the sensitivity of the interference effects to the size
of the source and the width of the frequency filter used for observations.
For a monochromatic
incoherent source with surface brightness $I(x_{\rm s},y_{\rm s})$ one finds
its flux by the convolution:
$$
F \sim \int dx_{\rm s} \int dy_{\rm s}~I(x_{\rm s},y_{\rm s})~|\Psi(x_{\rm
s},y_{\rm s})|^2
\eqno (13)
$$
In Figure 2 we show light curves of monochromatic sources crossing the
caustic. All sources are of the gaussian shape with characteristic sizes
of $0.1$, $0.3$, $1$, $3$ and $10 \times d_{\rm f} $.
For sources with sizes $ d_{\rm s} < d_{\rm f} $ the interference patern is
clearly visible
and present at any distance from the caustic. The asymptotic expansion
of Airy function (Abramovitz \& Stegun 1964)
${\rm Ai}(-x) \sim \sin(x^{3/2}+\pi/4)$
where we neglected the slowly changing amplitude, shows, that the spatial
distance between consecutive fringes is
$$
\Delta r_{\rm s} = {2\pi \over 3}~ d_{\rm f} ~ \left( {d_{\rm f} \over r_{\rm
s}} \right) ^{1/2}
% = { \pi \over 3 } ~ { d_{\rm F} \over 2^{1/3} } ~
% \left( {d_f \over r_{\rm s}} \right) ^{1/2}
% \left( {d_F \over d_{\rm c}} \right) ^{1/3} ~~.
\eqno (14)
$$
The spacing $ \Delta r_{\rm s} $ is proportional to the smoothed magnification,
$ A \sim \left( d_{\rm f} / r_{\rm s} \right) ^{1/2} $.
For a source moving with the velocity $V$ as measured in projection onto
the lens plane and perpendicular to the caustic, the characteristic time
for time variations is given as
$$
\Delta t \equiv {\Delta r_{\rm s} \over V }
\eqno (15)
$$
This spatial and temporal characteristics of interference fringes is
a generic feature of a motion across any caustic.

Let us consider now a typical photometric measurement using a broad band
filter with a band pass
$\Delta \omega/\omega =0.2$ We assume the source spectrum to be flat
in the filter range and use the "top hat" filter. We make convolution
in frequency  as well as in space. (Amplitude $\Psi$ depends on frequency
through $d_{\rm F}$ and $d_{\rm f}$). The results are shown on Figure 3
for the same shapes of sources as previously. Only close to the caustic
(where the phase of $\Psi$ is the same for all frequencies) the
interference pattern can be noticed. Farther away the fringes disappear.
The number of clear fringes is $\sim \omega/\Delta\omega$.

\vskip 0.5cm
\centerline{\bf 3. APPLICATION}
\vskip 0.5cm

The caustic patterns were modeled by Wambsganss (1990),
Witt (1990), Wambsganss, Witt, \& Schneider (1992), and Witt et al.
(1993, hereafter WKR) among
others. The last of the quoted papers can be applied directly to our
problem, since it gives the useful characteristics of caustics in the
lensing system of Q2237+0305. We use the image A characteristics in our
parameter estimation. The dependence of source magnification near
caustic on the average microlens mass given by WKR shows
that the relevant scale is the Einstein radius calculated for the
average microlens mass:
$$
r_{\rm E}=\left( {4GM_{\rm av}D \over c^2} \right) ^{1/2}
=1.6 \times 10^{16}{\rm cm}~
\left( M_{\rm av}/M_{\odot} \right) ^{1/2}~h_{75}^{-1/2}~~,
\eqno (16)
$$
where $h_{75}$ is the Hubble
constant in units of $75~{\rm km}~{\rm s}^{-1}~{\rm Mpc}^{-1}$.
Now using WKR Table 4 and translating their notation to ours we get the
value of
$\alpha \equiv \langle K^2\rangle/\langle m\rangle$
where $K$ is their flux factor and $\langle m \rangle$
is the averaged microlens mass in solar units. For image A we
get $\alpha=0.8$.
%MJ: wyrzucilem ten fragment
% nie rozumiem co to znaczy: "smoothly distributed mass component"
% tak czy owak chodzi zapewne o "macro-lens" i "macro-image" ?
The value of $ \alpha $ may be somewhat different for other macro-image
parameters.
Substituting we have:
$$
d_{\rm c} = \alpha~r_{\rm E} = 1.28 \times  10^{16}{\rm cm}~
\left( M_{\rm av}/M_{\odot} \right) ^{1/2}~h_{75}^{-1/2}~~.
\eqno (17)
$$
The Fresnel length for an optical wavelength calculated for the same
system is
$$
d_{\rm F}=6 \times 10^{10}{\rm cm}~\lambda_{5000}^{1/2}~h_{75}^{-1/2}~~,
\eqno (18)
$$
where $\lambda_{5000}$ is the wavelength in units of $5000~{\rm \AA}$.
Using this estimates of the characteristic scales we find that the
limiting size for a source to show the diffraction effects
at optical wavelength is:
$$
\hat{d}_{\rm s} < \hat{d}_{\rm f}
= {D_{\rm s} \over D_{\rm g}}~d_{\rm f} \approx
3 \times 10^{9}{\rm cm}~
\left( M_{\rm av}/M_{\odot} \right) ^{-1/6}
{}~\lambda_{5000}^{2/3}~h_{75}^{-1/2}~~,
\eqno (19)
$$
This is the upper limit to the source size
as measured in the source plane, which for
Q2237+0305 is at 8 times the distance to the lens plane.
Also, this is the best resolution of a quasar that can be achieved
through gravitational microlensing.
% The source must also be smaller than the intercaustic separation
% $d_{\rm diff}  \le r_{\rm E}~D_{\rm s}/{\rm D}_{\rm g}$ -- otherwise
% the effects related to different caustics would appear simultaneously
% and the generic characteristic of the fringes would be lost.
% Since the Einstein ring radius $r_{\rm E}$ and the characteristic caustic
% scale $d_{\rm c}$ are not different by many orders of magnitude, one can say,
% that the source must be always smaller than the Fresnel length projected into
% the source plane.
% Using both criteria and changing the masses of lenses one can find
% an absolute upper limit to the source size, which would be capable
% of producing interference effects. For the lensing system parameters
% of Q2237+0305 that is:
% $$
% d_{\rm diff} \le 4.7 \times 10^{11}{\rm cm}~\lambda_{5000}~h_{75}^{-1/2}~~
% \eqno (20)
% $$
% and the maximum is attained for a typical mass $ M =1.4 \times 10^{-11}
% ~ M_{\odot} ~ \lambda_{5000} $
% (or about $3 \times 10^{22}{\rm g}$ for optical wavelengths).

A characteristic time of femtolensing variability (assuming the relative
velocity of $600 {\rm km}~{\rm s}^{-1}$ at the lens plane which translates
to $5000 {\rm km}~{\rm s}^{-1}$ at the source plane, values used
in Q2237+0305 modeling) is
$$
\Delta t \approx 6~{\rm s}~
\left( M_{\rm av}/M_{\odot} \right)
^{-1/6}~\lambda_{5000}^{2/3}~h_{75}^{-1/2}~~.
\eqno (20)
$$

\vskip 0.5cm
\centerline{\bf 4. DISCUSSION}
\vskip 0.5cm

In our calculations we neglected the flux coming to an observer from
images not related to the critical line.  This is justified
as the source crossing of a caustic
does not influence other images and the pair of merging images
dominates the total flux.

The diffraction limits the resolution achievable through
photometric monitoring of caustic crossing to
$ \hat{d}_{\rm f} \approx 3 \times 10^{9} {\rm cm} ~
\left( M_{\rm av}/M_{\odot}
\right) ^{-1/6} ~ \lambda_{5000}^{2/3}~h_{75}^{-1/2} $
where $ M_{\rm av} $ is the average
mass of microlensing objects in the galaxy that macrolenses Q2237+0305.
This corresponds to the angular resolution of
$ \sim \hat{d}_{\rm f}/D_{\rm s} $, where $ D_{\rm s} $ is the
angular diameter distance to the source,
i.e. the resolving power of a caustic crossing in the Huchra's lens
is $ \sim 10^{-18} $ radians or 0.2 pico arcseconds in optical
light for stellar mass microlenses.  This is a very impressive
resolving power indeed.

If there is a substructure in the Q2237+0305 with a scale smaller
than $ \hat{d}_{\rm f} $ then interference pattern should be detectable
photometrically during the caustic crossing events:
a series of luminosity fluctuations with increasing or decreasing
amplitude with the characteristic time scale of few seconds (cf. eq. [20]).
Notice that if such fluctuations were detected they would allow the
determination of the transverse velocity of the caustic, or equivalently
they would allow the determination of a relation between the time scale
of light variability and the linear size at the source using single
site observation.  If no interference effects are detected the determination
of the caustic velocity requires at least two observing sites separated
by about one astronomical unit, as first pointed out by Grieger et al (1988).

Our estimates demonstrate that in the case of Q2237+0305, and similarly other
quasars, the geometrical optics is a very accurate approach. Investigating
light curves of quasars crossing caustics one can probe their spatial
structure down to scale as small as $ 0.05 R_{\odot} $.  Currently we
have no reason to expect that any quasar has structure as small as a tenth
of the solar radius, but it is good to know that structure down to such
a small scale can be resolved with gravitational microlensing.
It is also important to realize that the variation of image brightness
should be monitored on the shortest time scales allowed by
current technology in order to determine observationally what is the
highest magnification present at caustic crossings, and what is the
shortest time scale on which the brightness varies.

Adopting the geometric optics approximation we can estimate the source size
$ d_{\rm s} $ combining the equations (1b) and (2):
$$
d_{\rm s} = { \alpha \over \left( A_{\rm max} - A_{\rm min} \right) ^2 } ~
r_{\rm E } , \eqno(21)
$$
where $ \alpha \approx 1 $ (cf. Table 4 of WKR), $ r_{\rm E} $
is the Einstein ring radius of the microlensing masses, $ A_{\rm max} $
is the maximum magnification reached at the caustic crossing, and
$ A_{\rm min} $ is the magnification just outside the caustic.
Notice, that the estimate does not require any knowledge of either
the transverse velocity $ V $, or the angle $ \theta $ (cf. eq. [1a]).

We thank Hans Witt for the discussion.
This project was supported with the NSF grant AST 93-13620, NASA grant
NAG5-2759 and KBN grant 2-P03D-020-08.

\vskip 0.5cm
\centerline{\bf REFERENCES}
\vskip 0.3cm

\wc{Abramovitz, M. \& Stegun, I. A. 1964, {\it Handbook of Mathematical
	    Functions} (Washington: U.S.Gov. Printing Office)}

\wc{Blandford, R. D. \& Narayan, R. 1986, ApJ, 310, 568}

\wc{Czerny, B., Jaroszy\'nski, M., \& Czerny, M. 1994, MNRAS, 268, 135}

\wc{Deguchi, S. \& Watson, W.D. 1986, ApJ, 307, 30}

\wc{Gould, A. 1992, ApJ, 386, L5}

\wc{Grieger, B., Kayser, R., \& Refsdal, S. 1988, A\&A, 194, 54}

\wc{Huchra, J., Gorenstein, M., Kent, S., Shapiro, I., Smith, G.,
	   Horine, E., \& Perley, R. 1985, AJ, 90, 691}

\wc{Jaroszy\'nski, M. \& Marck, J.-A. 1994, A\&A, 291, 731}

\wc{Jaroszy\'nski, M., Wambsganss, J., \& Paczy\'nski, B. 1992,
		ApJ, 396, L65}

\wc{Kayser, R. \& Witt, H. J. 1989, A\&A, 221, 1}

\wc{Kundi{\' c}, T. \& Wambsganss, J. 1993, ApJ, 404, 455}

\wc{Kundi{\' c}, T., Witt, H. J., \& Chang, K. 1993, ApJ, 409, 537}

\wc{Peterson, J. B. \& Falk, T. 1991, ApJ, 374, L5}

\wc{Pen, U. E. et al. 1994, {\it Proceedings of the 31-st Liege
	  International Astrophysical Colloquium} "Gravitational Lenses
	  in the Universe", edited by Surdey, J., Fraipont-Caro, D.,
	  Gosset, E. \& Remy, M., Universite de Liege, p. 111}

\wc{Racine, R. 1992, ApJ, 395, L65}

\wc{Rauch, K. P. \& Blandford, R. D. 1991, ApJ, 381, L39}

\wc{Schneider, D. P., Turner, E. L., Gunn, J. E., Hewitt, J. N.,
 	Schmidt, M., \& Lawrence, C. R. 1988, AJ, 95, 1619}

\wc{Schneider, P. \& Schmidt-Burgk, J. 1985, A\&A, 148, 369}

\wc{Schneider, P. \& Weiss, A. 1986, A\&A, 164, 237}

\wc{Stanek, K.Z., Paczy\'nski, B., \& Goodman, J. 1993, ApJ, 413, L7}

\wc{Ulmer, A. \& Goodman, J. 1995, ApJ, (in press);
	also: Princeton Observatory Preprint No 569 (UG)}

\wc{Wambsganss, J. 1990, PhD Thesis, Ludwig-Maximilians University,
	Munich}

\wc{Wambsganss, J. 1992, ApJ, 392, 424}

\wc{Wambsganss, J. \& Paczy\'nski, B. 1991, AJ, 102, 864}

\wc{Wambsganss, J. \& Paczy\'nski, B. 1994, AJ, 108, 1156}

\wc{Wambsganss, J., Paczy\'nski, B., \& Schneider, P. 1990, ApJ,
		352, 407}

\wc{Wambsganss, J., Witt, H. J., \& Schneider, P. 1992, A\&A, 258, 591}

\wc{Witt, H. J. 1990, A\&A, 236, 311}

\wc{Witt, H. J., Kayser, R., \& Refsdal, S. 1993, A\&A, 268, 501 (WKR)}

\wc{Witt, H. J. \& Mao S. 1994, ApJ, 429, 66}

\vfill

\eject

\centerline{\bf FIGURE CAPTIONS}

\wc{Fig.1.  A schematic representation of a generic light curve
        when a point source crosses a caustic, as expected
         within a geometric optics approximation.}

\wc{Fig.2. Caustic-related magnification for gaussian shaped,
	monochromatic sources as a function of their distance from the
	caustic.  The plots are for the sources
	with the characteristic size of $0.1$ (the thinnest solid line),
	$0.3$, $1$, $3$  and $10 \times d_{\rm f}$ (the thickest line),
	where $d_{\rm f}$ is characteristic length near the caustic (eq. [12]).
	The geometrical optics result for a point source (eq. [6]) is
	shown for comparison as a dotted line.
	We adopt parameters of Q2237+0305 lensing system (eqs. [17,18])
	for $M_{\rm av}=1~M_\odot$ and $\lambda_{5000}=1$ to get
	the magnification values near the caustic.}

\wc{Fig.3. The same as on Fig.2 but for sources seen through a broad filter
	$\Delta \omega/\omega=0.2$.}

\vfill
\eject
\end

\begin{figure}[t]
\plotfiddle{fig1.ps}{8cm}{0}{50}{50}{-160}{-90}
\caption{Light curves for gaussian shaped sources crossing the caustic
	as seen in a single frequency. The plots are for the sources
	with the characteristic size of $0.1$ (the thinnest solid line),
	$0.3$, $1$, $3$  and $10 \times d$ (the thickest line).
	"$d$" is characteristic length near the caustic - see the text.
	The curves are parametrized by the distance from caustic $r_{\rm s}$
	rather than time and the units of flux are arbitrary.
	Only the flux from caustic-related images is included. In more
	realistic situations one should add a slowly varying flux from
	other images, but close to the caustic this part should be small.}
\end{figure}

\begin{figure}[t]
\plotfiddle{fig3.ps}{8cm}{0}{50}{50}{-160}{-90}
\caption{The same as on Fig.2 but for sources seen through a broad filter
	$\Delta \omega/\omega=0.2$.
}
\end{figure}

\end{document}